\documentstyle[preprint,eqsecnum,aps]{revtex}

\begin{document}
\draft
\preprint{HEP/123-qed}
\title{Nontopological thermal solitons in isotropic ferromagnetic lattices}
\author{N. Theodorakopoulos}
\address{
Theoretical and Physical Chemistry Institute\\
National Hellenic Research Foundation\\
Vas. Constantinou 48, GR - 116 35 ATHENS, Greece}
\maketitle
\begin{abstract}
The paper deals with the properties of thermally excited solitons of the
isotropic spin-$S$ ferromagnetic chain with
nearest-neighbor logarithmic interactions.
The exact statistical mechanics of the interacting soliton gas
is developed for the general case (arbitrary $S$, temperature and magnetic
field). At low temperatures the model's thermodynamics coincides with that of
the Heisenberg model.
We present analytical approximations of the leading-order asymptotic behavior
of the energy in three limiting cases:
(a) zero field, low temperature, classical limit;
(b) zero field, $T\to 0$, $S$ finite (quantum limit);
(c) zero field, high temperature, classical limit.
Cases (a) and (c) are examples of a dense gas of [non-topological] solitons;
results are in agreement with those obtained by the transfer integral
method. Case (b) illustrates the behavior of a dilute, yet strongly interacting
soliton gas; results for the thermodynamics are very close to (but not
identical with) spin-wave and/or Bethe-{\it Ansatz} predictions.
\end{abstract}
\section{Introduction}
Theoretical and experimental work with one dimensional
magnets has provided
ample evidence for the existence and importance of solitons\cite{MiSt91}.
The development of solitonic concepts and techniques has been characterized by
a remarkable degree of cross-fertilization between statistical physics,
field theory and applied mathematics.
Indeed, much of our present understanding of the thermal behavior
of magnetic solitons results from
Mikeska's seminal exploitation\cite{Mi78} of the formal equivalence
between the anisotropic Heisenberg model
(with an external field) and the Sine-Gordon continuum.
There has been no lack of \lq\lq direct\rq\rq\- candidates from magnetism.
The isotropic Heisenberg ferromagnet (IHF)
has been known to be integrable in the continuum
limit\cite{Takht77,Fog80},
and a variant, the Ishimori-Haldane-Fadeev ferromagnet (IHFF)
is completely integrable on a lattice \cite{Ishi,Hal82,Fa82}.
In both models, the properties of solitons at thermal equilibrium,
although of fundamental interest,
are difficult to extract. The reason for this is twofold:
(i) the presence of non-topological solitons, whose amplitudes can
become arbitrarily small and their numbers presumably arbitrarily large.
(ii) the fact that those non-topological solitons,
which perform breather-like internal oscillations in addition to
their translational motion, actually reduce to linear spin waves
in the small amplitude limit.
The first problem is generic to all systems which
carry non-topological solitons ({\it e.g.} the Toda lattice
\cite{Todabook}); it means that one must incorporate interactions between
solitons in order to achieve a valid  description of their statistics at
any finite temperature\cite{ThBa92}.
The second problem has been dealt with in the case of Sine-Gordon breathers.
There, it has been demonstrated (in leading order asymptotics \cite{Th84b})
that if one allows the presence of both breathers and phonons,
enough of either will be displaced to maintain the total number
of degrees of freedom.
Double counting can in fact be avoided less tediously {\it a priori}:
exact formulations of statistical mechanics in terms of {\it either}
nonlinear phonons\cite{Bull86} {\it or} breathers\cite{Fow86} yield
identical results.\par
The present work deals with the thermodynamics of IHFF
solitons. At sufficiently low temperatures, the model's properties
coincide with those of the IHF. Since the [integrable] IHF continuum can
be considered to be a good approximation to the IHF lattice
under the same condition, {\it i.e.,} at low temperatures, our results
in effect provide a description of soliton statistical mechanics
of the IHF as well. This equivalence has been exploited in a recent
Letter \cite{ThBa91}, where many of our key results have appeared
without proof\cite{Th_unpub91}. \par
The aims of this paper are: (i) to give a reasonably complete account
of the theory of [exact] soliton statistical mechanics in the only
magnetic system where this has been possible to a full extent and,
(ii) to support the conclusions of the theory by presenting its
application to selected analytically tractable limiting cases.
Of particular interest in the latter context is the treatment of
the high-temperature regime (not reported in Ref. \cite{ThBa91}), which
validates the concept of a strongly interacting, dense gas of
extremely localized solitons.\par
The paper is organized as follows: Section II introduces the model
and its dynamics. It reviews the basic properties of one- and
two-soliton solutions and discusses the conservation laws
which control the symmetries of soliton/soliton phase
shifts and are thus central to the statistical
mechanics. Section III formulates the statistical mechanics in terms of
occupation probabilities for microstates. Its main result is formulated in
terms of a two-dimensional integral equation. It will be shown that the exact
thermodynamic quantities (soliton density, energy, magnetization)
are controlled by
the asymptotic behavior of soliton quasienergies in phase space.
Section IV deals with the limiting case of zero-field,
high and low-temperature asymptotic behavior. In the latter case,
it will be possible to derive results for both the quantum and the
classical regime. Concluding remarks are made in Section V.
The relationship between thermodynamic properties
and integral-equation asymptotics is derived in the Appendix.
\section{Dynamics}
\subsection{The model}
The IHFF model Hamiltonian
\begin{equation}
H = -2JS^{2}\sum_{n} \ln \bigl(\frac{1}{2}+
    \frac{\vec{S}_{n}\cdot\vec{S}_{n+1}}
                        {2S^{2}}\bigr)
   - g B \sum_{n}({S}_{n}^{z}-S) \> ,
\label{Ham}
\end{equation}
describes a chain of $N$ atoms, each of which carries a spin ${\vec{S}_{n}}$
of length $S$,
and is subjected to an external field
$B$ along the z-axis. The particular form of the interaction
guarantees complete integrability
\cite{Ishi,Hal82,Fa82} at the classical limit,
{\it i.e.}, as $ S \to \infty, \hbar \to 0,
\hbar S \to S_{c}, J \to 0,  J S^{2} \to j, g \to 0, gS \to g_{c} $
\cite{Fish64}.
It will, however, be necessary to keep the value of $S$ as a parameter in
the problem, both in order to deal with semiclassical approximations
of quantum chains and because the equation which describes
statistical mechanics at the classical limit is singular.   \par
The classical dynamics of (\ref{Ham}) is described by
\begin{equation}
\frac{d}{dt} \vec{S}_{n} = \frac {\partial H}{\partial \vec{S}_{n}}
                  \times   \vec{S}_{n}
\label{speqmotion}
\end{equation}
where time is measured in units of $\hbar/(JS)=S_{c}/j$.
The fact that the dynamical system (\ref{speqmotion})
is completely integrable has been demonstrated in a variety of ways.
Ref.\cite{Ishi} makes use of the gauge equivalence
of the dynamics arising from (\ref{speqmotion}) (for $B=0$)
to the Ablowitz-Ladik(AL)\cite{AL76}
version of the discrete nonlinear Schr\"odinger
equation.
Refs.\cite{Hal82,Fa82} conjecture the existence of such a model by
considering a sequence of finite-$S$ models which are Bethe-Ansatz
solvable \cite{BaTa82}. The complete
inverse scattering theory has also been presented\cite{Papa87}.
We will follow \cite{Ishi} and make direct reference to
\cite{AL76} when appropriate.\par
\subsection{Properties of single solitons}
The problem of the magnetic field is disposed of first. The physical
effect of the field is to make all spins perform a
uniform precessional motion with angular frequency
$h=g_{c}B/j$.
This motion can be eliminated by introducing
a rotating coordinate frame. Now, since the dynamics of the
zero-field IHFF (\ref{speqmotion}) can be mapped  to the the dynamics of
the AL model, it is possible to describe any general multi-soliton
solution of (\ref{speqmotion}) in terms of the corresponding AL solution,
and superimpose the uniform precession due to the magnetic
field at the end. This simplification is of particular importance for
the statistical mechanics, since it means that all relevant properties
of interactions between solitons are controlled by the values of their
dynamical parameters for vanishing magnetic field; it does not imply
however that thermodynamic properties have a trivial dependence on
the magnetic field ({\it cf.} Section III). \par
The form of the single soliton solution ({\it cf. }\cite{Ishi}),
demonstrates that all physical soliton properties can be described
in terms of two parameters, $w$ and $k$. Thus, the soliton has a
spatial extent (in units of the lattice constant
${\ell}$) equal to $1/(2w)$, it moves with a
translational velocity
$ U(w,k) = \sinh 2w \sin 2k /(2w)$,
and performs an internal oscillation of frequency
$ \Omega(w,k) = 2kU(w,k) + 2(\cosh 2w \cos 2k - 1) + h  $;
furthermore,
one can evaluate the two obvious integrals of motion, namely, the
total magnetization
\begin{equation}
M(w,k) = - 2  \> \frac{\sinh 2w}{\cosh 2w - \cos 2k}  ,
\label{M1}
\end{equation}
(in units of $S_{c}$),
and energy
\begin{equation}
E(w,k) = 8  w - h M(w,k)
\label{E1}
\end{equation}
(in units of $j$) carried by a single soliton.
The soliton canonical momentum $P$,
defined\cite{Th88} via
$ \left(\partial P/ \partial E\right)_{M} =   1 /U  $,
(in units of $S_{c}/{\ell}$),
determines the correct measure to be used in statistical mechanics:
\begin{equation}
        dP dM = \frac{\partial(P,M)}{\partial(w,k)} dw dk\\
        = \frac{8}{U} \left( \frac{\partial M}{\partial k} \right)_{w}
dw dk   =32 \frac{w}{(\cosh 2w -\cos 2k)^{2}} dw dk
\label{measure}
\end{equation}
In the limit $w\to 0$, the properties of a soliton characterized by $(w,k)$
reduce to those of a plane spin wave with wavevector $q=2k$ and frequency
$4\sin^{2}(q/2)$.\par
\subsection{Multisoliton solutions: soliton-soliton phase shifts}
The simplest way to evaluate  the asymptotic space shifts resulting from
the interaction of two solitons is to go back  to
the AL formulation of the inverse-scattering transform,
write down the two-soliton solution, and examine its
asymptotic form as  $t \to \pm\infty$. The calculation is tedious but
straightforward.
Here, we only quote the result and make a few comments. The phase shift
which a soliton  $(w,k)$ experiences due to its collision with
a soliton $(w',k')$ is given by
\begin{equation}
\Delta(w,k;w',k') = \frac{1}{2w} \> \ln  \Biggl[ \frac
{\cosh 2(w'+w)-\cos 2(k-k')}
{\cosh 2(w'-w)-\cos 2(k-k')} \Biggr] \>.
\label{ss}
\end{equation}
\indent
It follows that the sum rule
\begin{equation}
\int_{0}^{\pi}  dk  \> \Delta(w,k;w',k')=
\frac{2 \pi}{w} \min(w,w')
\label{sumrule}
\end{equation}
is always satisfied. Furthermore, the  soliton/soliton
phase shifts obey the following symmetry (generalizable to
any multisoliton solution characterized by
$\{ w_{\lambda},k_{\lambda} \}$):\par
\begin{equation}
\sum_{\lambda \neq \lambda '} w_{\lambda}
\Delta(w_{\lambda},k_{\lambda};w_{\lambda '},k_{\lambda '})= 0
\label{magsym}
\end{equation}
\subsection{Phase shift symmetries and conservation laws}
Symmetries of the type (\ref{magsym}) are generic to
soliton-bearing systems.
They reflect the existence of a low-order integral of motion
$I = \sum_{\lambda}w_{\lambda}U_{\lambda}$   ,
(where $ U_{\lambda}$ is the velocity of the $\lambda th$ soliton)
and imply that the [conserved] quantity  $ \sum_{\lambda}w_{\lambda}$
travels at a constant speed.
Inspection of the formula for the group velocity shows
that the required conserved quantity is AL's $C_{1}$,
or, in the spin language,
\begin{equation}
    I = \sum_{n} {\tilde\chi}_{n} \nonumber \\
     \equiv\sum_{n}  \frac{\vec{S}_{n}\cdot(\vec{S}_{n+1} \times
     \vec{S}_{n-1})}
        {(1+\vec{S}_{n}\cdot\vec{S}_{n+1})
        (1+\vec{S}_{n}\cdot\vec{S}_{n-1})} ,
\end{equation}
and reflects  the invariance of the original spin Hamiltonian
under infinitesimal rotations of each spin $\vec{S}_{n}$ around
an axis defined by the spin-space gradient
\begin{equation}
\nabla_{\vec{S}_{n}}I
\equiv \nabla_{\vec{S}_{n}} ( {\tilde\chi}_{n+1} + {\tilde\chi}_{n}
+ {\tilde\chi}_{n-1} )     \> .
\label{rot}
\end{equation}
Note that generally ({\it i.e.} for $h \neq 0$) $w$ is proportional to
the exchange part of soliton energy. It is possible to
show that the center of exchange energy of the entire IHFF
lattice travels at a constant speed, independently of
any soliton considerations. Given an
infinite chain with
any type of isotropic nearest neighbor
spin interaction $f(\vec{S}_{n}\cdot\vec{S}_{n+1})$,
consider the quantity
\begin{equation}
        X = \frac{1}{Z}\sum_{n}
   n \> g(\vec{S}_{n}\cdot\vec{S}_{n+1})  \>,
\label{X}
\end{equation}
where $g$ is such that the denominator $Z = \sum_{n}
[ g(\vec{S}_{n}\cdot\vec{S}_{n+1}) ] $
is a constant of the motion. The derivative $dX/dt$  can be written
down by using (\ref{speqmotion});
it turns out that, in order for it to remain constant
under the Hamiltonian flow,  (i)
$g$ must be proportional to $f$ (and thus $Z$
proportional to the exchange part of the Hamiltonian)
and, (ii) the quantity
\begin{equation}
\sum_{n} g'(\vec{S}_{n}\cdot\vec{S}_{n+1})
       g'(\vec{S}_{n}\cdot\vec{S}_{n-1})
   \vec{S}_{n}\cdot(\vec{S}_{n+1} \times  \vec{S}_{n-1})
\label{I3}
\end{equation}
must be a constant of the motion, as indeed is the case with the
logarithmic (but not with the Heisenberg) interaction.
Thus, the existence of the constant of motion $I$ guarantees a
very special status to the exchange energy of the IHFF model (analogous
to that of the mass in a lattice model\cite{Todabook}).
\section{Soliton thermodynamics: exact results}
The general part of the formalism used has been presented in \cite{ThWe},
in the context of the
Sine-Gordon equation;  the reader is referred there
for details.
The physics involve the minimization of
the classical free energy functional with respect to the
occupation probability of each state. The
procedure is self-consistent since the density of states
is itself a functional of the occupation probabilities, due to
the pairwise additive interactions (\ref{ss}) which restrict available
phase space.
The thermodynamics derived within this framework is
in principle exact, with a single caveat: because solitons with a very
large characteristic length are in practically indistinguishable from linear
excitations (spin waves, {\it cf.} above) one expects that a correct
statistical mechanical treatment
will ultimately eliminate one or the other. The situation is analogous
with the famous case of the Sine-Gordon breather. It appears therefore
expedient to formulate the theory in terms of solitons alone
({\it cf.} \cite{Sa86}).\par
The thermodynamic potential (free energy)
of the interacting soliton gas is given by
\begin{equation}
-\beta F   =    \int d\Gamma R_{0}(\Gamma) \bar{n}(\Gamma)
\label{F}
\end{equation}
where $\beta=JS^{2}/k_{B}T$ is the dimensionless inverse temperature,
$\Gamma $ is shorthand for the point $(w,k)$ in the soliton parameter
phase space, and the phase space element $d\Gamma$ is defined by the measure
(\ref{measure}).
The density of states $R_{0}(\Gamma)$ refers to the non-interacting soliton
gas (with our choice of units ({\it cf.} Section IIb)
equal to $N/(2\pi)S^{2}$)
and the occupation probability $\bar{n}(\Gamma ) \equiv \exp(-\beta
\epsilon)$
defines a characteristic [field- and temperature-dependent] quasiparticle
energy $\epsilon (\Gamma)$.
The latter is determined by the integral equation
\begin{equation}
\beta \epsilon(\Gamma) = \beta E(\Gamma) +
\frac{1}{2\pi}\int d\Gamma '\Delta (\Gamma ',\Gamma)
e^{-\beta\epsilon(\Gamma ')}   \>\>      .
\label{quasien}
\end{equation}
A further quantity of interest is the density of available
states at any given point in phase space
expressed as fraction of $R_{0}(\Gamma)$, {\it i.e.,}
$     \hat{R}(\Gamma) \equiv R(\Gamma)/R_{0}(\Gamma) $.
In the IHFF case, equations (\ref{F}) - (\ref{quasien}) take the form
\begin{equation}
-\beta f  =  \frac{16}{\pi}  \int_{0}^{\pi}dk \int_{0}^{\infty}dw
       \frac{w}{(\cosh 2w - \cos 2k)^{2}} e^{-\beta \epsilon (w,k)}  ,
\label{fIshi}
\end{equation}
and
\begin{eqnarray}
\beta \epsilon (w,k) = \beta E(w,k) +
      \frac{8}{\pi} S^{2} \int_{0}^{\pi}dk \int_{0}^{\infty}dw'
        &   \frac{1}{(\cosh 2w' - \cos 2k')^{2}}
 \nonumber    \\
        & \ln  \left[ \frac{\cosh 2(w'+w)-\cos 2(k-k')}
                {\cosh 2(w'-w)-\cos 2(k-k')} \right]
        e^{-\beta \epsilon (w',k')}       ,
\label{eIshi}
\end{eqnarray}
respectively, where $f$ is now the free energy per site. Furthermore,
\begin{equation}
\hat{R}(w,k) = \frac{1}{E_{0}(w)}
 \left( \frac{\partial \beta \epsilon (w,k)}{\partial \beta} \right)
   _{\beta h}              .
\label{dosIshi}
\end{equation}
The derivation is straightforward, and involves
use of (\ref{measure}) and (\ref{ss}).
Some special attention must be paid to the analogy
of (\ref{dosIshi}) with Eq.(19) of \cite{ThWe}; in this case,
the crucial symmetry refers to the exchange part of
the energy of a free soliton ({\it cf.} above); hence
the denominator
$E_{0}(w)=8w$ in (\ref{dosIshi}). Moreover, the derivative with respect
to $\beta$ must be taken at constant $ \beta h$. \par
The thermodynamic functions we are interested in, i.e. energy,
soliton density and magnetization can be obtained by differentiation
of (\ref{fIshi}) with respect to $\beta$, $\mu$ and $h$ respectively.
The soliton chemical potential must vanish, since the soliton density
is not {\it a priori} fixed. The details are presented in the Appendix.
The important feature is that we do not require detailed knowledge
of the solution of (\ref{eIshi}) at all points in the two-dimensional
region defined by  $ 0 < w < \infty$, $ 0 < k < \pi$. {\em
The thermodynamic behavior of the soliton gas is controlled
by the asymptotic properties of (\ref{eIshi}) in the limits}
$w\to 0$ {\em and} $w\to \infty$. In particular, the free energy
is given by
\begin{equation}
        \beta f   = -\frac{1}{2\pi}\int_{0}^{\pi} dk
                C_{\infty}(k)
\label{fIasy}
\end{equation}
and the soliton density per site by
\begin{equation}
        n_{s}   = \frac{1}{2\pi}\int_{0}^{\pi} dk
  \left[1 - \left( \frac{\partial V_{0} } {\partial V_{\infty}}
                        \right)_{\beta h}   \right]
\label{nIasy}
\end{equation}
where
\begin{mathletters}
\begin{equation}
V(w,k)  \equiv \frac{\partial \beta \epsilon(w,k)}{\partial w},
\label{Vdef}
\end{equation}
\begin{equation}
V_{\infty}   \equiv \lim_{w\to\infty} \> V(w,k)
   =  8\beta   ,
\label{vinfty}\
\end{equation}
\begin{equation}
V_{0}  \equiv \lim_{w\to 0} V(w,k),   {\rm and}
\label{v0}
\end{equation}
\begin{equation}
C_{\infty}(k)  \equiv  \lim_{w\to\infty}
\left( \beta \epsilon(w,k)-\beta E(w,k)
\right) .
\label{Cinfty}
\end{equation}
\end{mathletters}
It follows that the internal energy $u$ and magnetization $m$ per site
can be obtained by
direct differentiation of (\ref{fIasy})
with respect to $\beta$ and $h$ respectively,
${\it i.e.,}$
\begin{equation}
 u = \left( \frac{\partial \beta f}{\partial \beta}\right)_{h}
   =  -\frac{1}{2\pi}\int_{0}^{\pi} dk
     \left(     \frac{\partial  C_{\infty} (k) }
            {\partial \beta }\right)_{h}
\label{uIasy}
\end{equation}
\begin{equation}
m = \left( \frac{\partial \beta f}{\partial h}\right)_{\beta}
  =  -\frac{1}{2\pi}\int_{0}^{\pi} dk
    \left(     \frac{\partial  C_{\infty} (k) }
            {\partial h}\right)_{\beta}\> .
\label{mIasy}
\end{equation}
\section{Soliton thermodynamics: analytic approximations}
\subsection{The classical, zero-field, low temperature case}
At low temperatures we expect
thermodynamic properties to be dominated by solitons whose exchange energies
are lower than $1/\beta$, {\it i.e.,} with parameter values
$w\ll1$. If, in addition, $k\ge w$, it is possible to approximate
the exact phase shift (\ref{ss}) by
\begin{equation}
\Delta(w,k;w',k') = \frac{2\pi}{w} \> \min(w,w') \> \delta (k-k')
\> \> \>.
\label{ssapp}
\end{equation}
Note that the above approximation preserves the symmetry (\ref{magsym}) of the
soliton/soliton phase shift
and satisfies the sum rule (\ref{sumrule}), so we can reasonably expect it
to exhibit at least the salient features of the
exact solution of (\ref{eIshi}).
Introducing (\ref{ssapp}) in (\ref{eIshi}), it is possible to decouple the
statistics of solitons with different $k-$values, and transform the
two-dimensional integral
equation (\ref{eIshi}) to a set of one-dimensional integral equations
in which $k$ appears as a parameter:
\begin{equation}
\beta \epsilon (w,k) = \beta E(w,k) +
             8 S^{2} \int_{0}^{\infty}dw'
       \frac{\min (w,w')}{(w'^{2} + \sin^{2}k)^{2}   }
        e^{-\beta \epsilon (w',k)}
\label{e1dIshi}
\end{equation}
The integral equation (\ref{e1dIshi}) can be transformed to a second-order
differential
equation for $x \equiv \beta \epsilon(w,k)$:
\begin{equation}
\frac{d^{2}x}{dt^{2}}
=  - 8 \> \frac{e^{-x-\beta h M(t,k)} }{  (1+ a t^{2} )^{2} } \> ,
\label{odeI}
\end{equation}
where $a=\sin^{2}k/S^{2}$ and we have used a rescaled independent
variable $t=S w/ \sin^{2}k$,
in terms of which  $M(t,k) = -2t/(1+ a t^{2} )/S$.
Eq. (\ref{odeI}) must be solved subject to
mixed initial and final conditions, {\it i.e.,} $x(0)=0, dx/dt(\infty)
=v_{\infty}=8( \beta /S )\sin^{2}k$. (Note the rescaling with respect to
$V_{\infty}$ of $(\ref{vinfty}))$\par
The remainder of this Section specializes to the case $h=0$.\par
In the classical limit, $S \to \infty, \beta   $\- finite,
the quantities
\begin{equation}
v_{\infty}=8\beta  \sin^{2}k /S
\label{vinfty_cl}
\end{equation}
and $a$ become infinitesimally small.
Eq.\ (\ref{odeI}) can be further simplified to
\begin{equation}
\frac{d^{2}x}{dt^{2}}
=  - 8 \> {e^{-x} }\> ,
\label{odeII}
\end{equation}
which, for the given boundary conditions,  has the exact solution
\begin{equation}
 x(t) = 2 \ln \left\{ \frac{4}{v_{\infty}}
   \sinh \left(  \frac{ v_{\infty}t}{2}
 + \sinh^{-1} \left(\frac{v_{\infty}}{4}\right)   \right)  \right\}.
\label{sol}
\end{equation}
Two remarks are in order here. First, that exactly the same equation
(\ref{odeII}) controls the classical limit of the Sine-Gordon
breather gas \cite{Sa86}. Second, that if one regards (\ref{odeI}),
with $h=0$, as a
mechanical system, (\ref{odeII}) represents its \lq\lq autonomous\rq\rq\-
limit.  The asymptotic properties of (\ref{sol}) are particularly simple
({\it cf.} definitions above):
\begin{equation}
v_{0} = \sqrt{16+v_{\infty}^{2}} \>,
\label{v0_aut}
\end{equation}
and
\begin{equation}
C_{\infty} = 2 \ln \left\{ \frac { v_{0} + v_{\infty}}{v_{\infty} }  \right\}.
\label{cinf_aut}
\end{equation}
It follows that
\begin{equation}
 - \frac{\partial C_{\infty}} { \partial v_{\infty} }
=   2 \left( 1 - \frac{\partial v_{0}} {\partial v_{\infty}} \right)  \>.
\label{dcdvinf_aut}
\end{equation}
In the classical limit under consideration,
This means ({\it cf.} Eq. (\ref{nIasy}))
that the magnetic lattice will be equally populated with
solitons of all $k$, producing a total density of $1/2$ solitons per site, as
expected from counting the classical degrees of freedom. The internal energy
density can be found by inserting (\ref{dcdvinf_aut}) and (\ref{vinfty_cl}) in
(\ref{uIasy}) to be equal to $k_{B}T$, as demanded by the equipartition
theorem. \par
\subsection{The finite $S$, zero-field, $\beta \to \infty$ limit}
The core of the approximation scheme developed in the preceding subsection
remains valid away
from the classical limit if the temperatures become extremely low.
More precisely, for $S$ finite and
$\beta$ large, consider the range of $k$ defined by
\begin{equation}
v_{\infty}  =  8 (\beta /S) \sin^{2}k  \le A  \>  ,
\label{vinfty_quantum}
\end{equation}
where $A\gg 1$ but finite. As $\beta \to \infty$, this range is bounded by
a characteristic $k_{c} \propto 1/\sqrt{\beta}$. For such $k$'s,
the parameter $a$ is smaller than $1/\beta$. In the limit $\beta \to \infty$
this  allows us to substitute
(\ref{odeI}) by (\ref{odeII}).
On the other hand, values of $k$ {\it outside} this range
are (in the limit $\beta \to \infty$) irrelevant. The argument for this is as
follows: such $k$-values lead to
$v_{\infty}\ge A \gg 1$;
numerically large values of
$v_{\infty}$ lead in turn to comparably large values of $v_{0}$,
 in the context of both (\ref{odeII})  and (\ref{odeI}).
In the former case this can be proved by
inspecting (\ref{v0_aut}). Moreover, it is in both cases clear  that a
$v_{\infty}\gg 1$ can only come from a $v_{0}\gg 1$. This however means that
$x(t)$ grows very rapidly from its initial value $x(0)=0$. As this happens,
the right hand sides of both (\ref{odeII}) and (\ref{odeI}) vanish, dominated
by the exponential factor. Since the  \lq\lq acceleration\rq\rq\- vanishes
before it has any time to act, the
\lq\lq velocity\rq\rq \- stays at its initial value, {\it i.e.,}
$v_{\infty} \approx v_{0}$,
\begin{equation}
1 -  \frac{dv_{0} } {dv_{\infty}}  \approx 0
\nonumber
\end{equation}
and therefore such solitons are not thermally excited ({\it cf.}(\ref{nIasy})).
As a consequence, in the limit $\beta \to \infty$, it is legitimate to use
(\ref{odeII})  in lieu
 of  (\ref{odeI}) {\it for all} $k$, and thus exploit the
properties  (\ref{v0_aut})-(\ref{dcdvinf_aut}) of the solution (\ref{sol}).
\par
Inserting (\ref{vinfty_quantum}) in (\ref{nIasy}), and
approximating $\sin k$ by its argument, we obtain the leading-order
contribution to the soliton density,
\begin{eqnarray}
n_{s}  & = & \frac{1}{\pi}
\sqrt{\frac{S}{2\beta}}
\int_{0}^{\infty} dx \left( 1 -  \frac{x^{2}} {\sqrt{1+x^{4}}}  \right)
\nonumber    \\
       & = & \frac{\sqrt{2\pi}}  {\left[ \Gamma ( \frac{1}{4}) \right]^{2}  }
           \left( \frac{k_{B}T} {JS} \right)^{1/2}    \> .
\label{n_s_quantum}
\end{eqnarray}
The corresponding result for the energy density, obtained by making use of
(\ref{vinfty_quantum}) in (\ref{dcdvinf_aut}), is
\begin{equation}
 u    = 2  \frac{\sqrt{2\pi}}{\left[ \Gamma ( \frac{1}{4}) \right]^{2}    }
          JS  \left( \frac{k_{B}T}{JS} \right)^{3/2}    \>.
\label{u_quantum}
\end{equation}
A few remarks are in order here. First, Eq. (\ref{n_s_quantum})
means that the
soliton gas is dilute. Not all solitons are excited. This is a characteristic
of the quantum regime. In the classical limit ({\it cf.} previous and following
subsection) solitons must exhaust all degrees of freedom. (It should be
remembered that no provision for linear excitations has been made.)
Second,  solitons
carry an average energy of $2k_{B}T$ each, as suggested by equipartition.
Numerical solution of (\ref{odeI}) indicates \cite{ThBa91} that this
result is only valid in the limit $\beta \to \infty$.
Third, the reason the approximation (\ref{ssapp}) works is that the conditions
for its validity are met by
thermal solitons;  characteristic $w$-values
of $1/\beta $ are much lower than the typical $k$ values of ${\cal O}(k_{c})$
(quantum case), or ${\cal O}(1)$ (classical case, preceding subsection).
Fourth, the numerical value of the prefactor of the energy, 0.3814,
is close to the spin-wave result of 0.3684. Comparable results have been
obtained in the $S=1/2$ case in terms of the Bethe-Ansatz
\cite{Schl85,Taka86}. In view of the
soliton-magnon duality at low amplitudes, this is not unexpected.
\subsection{The classical, zero-field, high temperature limit}
In the high temperature regime, $\beta < 1$, narrow solitons which
carry large amounts of exchange energy, {\it i.e.,} with values
of $w>1$ become predominant. As long as the characteristic parameters $w,w'$
of two such solitons do not come very close to each other,
the hyperbolic cosines in
the exact soliton/soliton phase shift (\ref{ss})
can be approximated by exponentials, leading to
\begin{equation}
\Delta(w,k;w',k') = \frac{2}{w} \> \min(w,w') \>   .
\label{ssapp2}
\end{equation}
The above approximation is the exact opposite of (\ref{ssapp}):
instead of a
$\delta$ function in $k$-space, which decouples soliton modes corresponding
to different $k$'s, (\ref{ssapp2}) is independent of $k,k'$, ensuring that
all soliton modes couple to each other.
It should be further noted that (\ref{ssapp2})
satisfies the sum rule (\ref{sumrule}), just like the low-$T$ approxination
(\ref{ssapp}).\par
Substitution of (\ref{ssapp2}) in (\ref{eIshi}) yields
\begin{equation}
\beta \epsilon (w,k) = 8 \beta  w+
        \frac{128 S^{2}}{\pi } \int_{0}^{\pi}dk'\int_{w_{0}}^{\infty}dw'
e^{- 4w'}\min (w,w')
        e^{-\beta \epsilon (w',k') }    \>,
\label{eIshiHT}
\end{equation}
where again the hyperbolic cosine in the denominator of
(\ref{eIshi}) has been approximated by an exponential, to
ensure consistency with (\ref{ssapp2}). It should be emphasized that these
approximations are crude, and that they cannot be expected to yield accurate
results for values of $w\ll 1$.  The point is that such low energy solitons are
irrelevant at high temperatures.\par
In accordance with this argument,
we temporarily exclude soliton modes with very low values of $w$
from any further consideration
by introducing a lower cutoff  in (\ref{eIshiHT}).  This will be useful in
controlling singularities, although the value of the cutoff will be set
to zero at the end of this calculation.\par
Since the right-hand side of (\ref{eIshiHT}) is $k$-independent, $\epsilon$
can only depend on $w$; Integration over $k'$ can be performed trivially, and
reduction to an o.d.e. follows the steps of the previous subsection.
The result is
\begin{equation}
\frac{d^{2}x}{dw^{2}}
   = -128 S^{2} e^{-(x+4w)}  \>  .
\label{odeIII}
\end{equation}
The most general solution of Eq. (\ref{odeIII}) consistent with
the condition $\lim_{w\to\infty}dx/dw \equiv V_{\infty} = 8\beta $ is
\begin{equation}
x(w) = 2 \ln \left\{ \frac{4}{\lambda}
      e^{-2w} S \sinh \left( 2\lambda w + b \right) \right\},
\label{solIII}
\end{equation}
where $\lambda = 1 + 2 \beta $.
We do not know the value of $b$, since we have no knowledge of the exact
behavior of $x(w) $as $w \to 0$.
The following procedure will be adopted instead:
$b$ will be set equal to zero, and the value of the cutoff will be fixed by
demanding $x(w_{0})=0$. This leads to $w_{0}=1/(8S)$, which vanishes
at the classical limit. The solution (\ref{solIII}) has the asymptotic
properties
\begin{equation}
V_{0}^{(S)} \equiv V(\frac{1}{8S}) = -4 + 16S \> \> \> ( S\gg1 )
\label{v0ht}
\end{equation}
and
\begin{equation}
C_{\infty} =  - 2 \ln \left( \frac {1+2\beta }{2} \right) + 2 \ln S \>.
\label{cinfht}
\end{equation}
Both (\ref{v0ht}) and (\ref{cinfht}) are valid for all $k$. In the limit
$S \to \infty$, $V_{0}^{(S)}$ diverges, but
$ \lim_{S \to \infty}   dV_{0}^{(S)}  / dV_{\infty} = 0$,
and thus, from (\ref{nIasy})  the soliton density must equal $1/2$.
Furthermore, from (\ref{uIasy}) and (\ref{cinfht}),
the internal energy per site
\begin{equation}
   u = \frac{2j}{1+2j/(k_{B}T)}
\label{uht}
\end{equation}
can be calculated, in agreement with the exact result obtained by the
transfer integral method
\cite {Web88}. \par
Although this completes the derivation of soliton thermodynamics at the
classical high temperature limit,
it is instructive to examine the basic features of the analytical solution
(\ref{solIII}). For example, using (\ref{solIII})
in the alternative definition of the soliton density
({\it cf.} Ref. \cite{ThWe}),
allows us to examine the $w$-distribution of thermal solitons (with the
understanding that no conclusions should be drawn about
details in the region $0<w\ll 1$):
\begin{equation}
n_{s} = 2 \lambda \int_{0}^{\infty} dw \frac{1}{\sinh^{2}(2\lambda w)}
        \left\{ \frac {2\lambda w} {\tanh (2\lambda w)}
                       - 1\right\}     .
\label{nhtint}
\end{equation}
Eq. (\ref{nhtint} ) implies that
solitons which are thermally excited have values of $w\approx 1/\lambda$.
In other words, typical thermal solitons cannot become narrower than a lattice
constant. Since each such soliton carries an energy of order $j$, the
total energy per site is of order unity at high
temperatures, as expressed by
(\ref{uht}).
\section{Concluding remarks}
We have presented an \lq\lq advanced phenomenology\rq\rq\- of the interacting
soliton gas for an isotropic ferromagnetic chain.
The theory is in principle exact but requires solving a two-dimensional
integral equation, {\it i.e.} in effect has the same degree of complexity as
the quantum mechanical Bethe-{\it Ansatz}.
In the case of vanishing magnetic field
it has been possible to develop valid approximations in the
high and low temperature regime and derive analytic expressions for
thermodynamic quantities. The results presented provide detailed information
about the manner in which
nonlinear modes accommodate themselves at finite temperatures.\par
In the low temperature regime, the fortunate occurrence of a
gapless spectrum allows a
semiquantitative understanding of the quantum, $T\to 0$, limit
of the Heisenberg chain with essentially classical techniques. The behavior of
the soliton density
suggests the presence of a \lq\lq dilute\rq\rq\- soliton gas.
The quotation marks imply that the term should be used with some caution. It
refers to typical distance between the centers of two solitons. At low
temperatures, this is indeed large ( of
order $T^{-1/2}$); however, the spatial extent of a thermal soliton is
of order $1/T$, {\it i.e.,} much larger. There is therefore significant
overlap between solitons, suggesting the presence of strong interactions.
It is interesting to digress at this point and consider the analogous situation
in the Toda lattice. There, the typical intersoliton distance at low
temperature is of the same order as the spatial extent of thermal solitons,
{\it i.e.,} $T^{-1/3}$. \par
It appears that a properly formulated soliton phenomenology has a range of
validity which far exceeds the limits imposed by the original
conceptual framework \cite{KruSch,Curr}. In the particular case of the
quantum low-temperature regime, marginal uncertainties in the
Bethe-{\it Ansatz} make it presently impossible to decide whether the soliton
result is asymptotically exact.
It is however clear, that soliton theory presents a better alternative to spin
waves: the magnetization obtained from the soliton
theory vanishes at any finite temperatures ({\it cf.} the divergent
magnetization obtained from spin-wave theory).\par
At high temperatures it appears, that in spite of the crudeness of our
approximation, the theory again describes leading order corrections. The
{\it exact} agreement of (\ref{uht}) with the transfer integral result should
be regarded as fortuitous. In this case the soliton gas is dense in the
traditional sense. Intersoliton distances and soliton spatial extent are both
in the range of the lattice constant.
\appendix
\section*{}
In order to derive (\ref{fIasy}), we subtract the \lq\lq bare\rq\rq\-
part of the
soliton energy, $\beta E(w,k)$ from both sides of (\ref{eIshi}) and integrate
both sides over all $k$. The result is
\begin{equation}
\int_{0}^{\pi} dk \left\{ \beta \epsilon - \beta E(w,k) \right\}
=       8 S^{2} \int_{0}^{\pi}dk \int_{0}^{\infty}dw'
       \frac{2}{(\cosh 2w' - \cos 2k')^{2}}        \\
\>      2 \min (w,w')
        e^{-\beta \epsilon (w',k')}   \>    ,
\label{fIasy_aux}
\end{equation}
where we have made use of the sum rule (\ref{sumrule}).
In the limit $w\to \infty$, the integrand of the left-hand side of
(\ref{fIasy_aux}) is $C_{\infty}$, whereas the right-hand side coincides,
except for a prefactor $1/(2\pi)$, with $-\beta f$ ({\it cf.}
Eq. \ref{fIshi}). This proves (\ref{fIasy}).\par
In order to derive (\ref{nIasy}), we differentiate both sides of
(\ref{eIshi}) with respect to $w$. In the limit $w\to 0$, this yields
\begin{equation}
V_{0}(k) = V_{\infty} +   \frac{4\beta h}{1-\cos 2k}
     +   \frac{      8 S^{2} }{\pi}
         \int_{0}^{\pi}dk \int_{0}^{\infty}dw'
       \frac{1}{(\cosh 2w' - \cos 2k')^{2}}
       \frac{4 \sinh 2w'}{\cosh 2w' - \cos 2(k'-k)}
        e^{-\beta \epsilon (w',k')}   \>    .
\label{V0_aux}
\end{equation}
Differentiating  both sides of (\ref{V0_aux}) with respect to
$V_{\infty}$ {\it at constant} $\beta h$, we obtain
\begin{equation}
\left( \frac{\partial V_{0}(k)}{\partial V_{\infty}}  \right)_{\beta h}
 =  1 -   \frac{      8 S^{2} }{\pi}
\int_{0}^{\pi}dk \int_{0}^{\infty}dw'
       \frac{8w'}{(\cosh 2w' - \cos 2k')^{2}}
       \frac{4 \sinh 2w'}{\cosh 2w' - \cos 2(k'-k)}      \\
\>      \hat{R}(w',k')e^{-\beta \epsilon (w',k')}   \>    ,
\label{V0_aux2}
\end{equation}
where we have used (\ref{dosIshi}) for the density of states $\hat{R}$.
Eq. \ref{nIasy} can now be obtained by integrating both sides
of (\ref{V0_aux2}) over $k$.\par
A further by-product of (\ref{V0_aux2}) is that, in the limit $k\to 0$, the
integrand in the right-hand side expresses the total magnetization in soliton
component form. This implies that the total magnetization per site (in units of
$S_{c}$) can be alternatively expressed in terms of the asymptotic properties
of (\ref{eIshi}), as
\begin{equation}
m = \lim_{k \to 0}
 \left( \frac{\partial V_{0}(k)}{\partial V_{\infty}}  \right)_{\beta h}  .
\label{mag2}
\end{equation}
The above alternative  can be useful in situations where the integral
(\ref{mIasy}) cannot be calculated exactly.
In all limiting situations which we have treated analytically ({\it cf.}
Section IV),
(\ref{mag2}) guarantees a vanishing magnetization.
\end{document}